\documentclass[12pt]{article}
\usepackage[english]{babel}
\usepackage{graphicx}
\usepackage{amsthm}
\usepackage{amssymb}
\usepackage{amsmath,amsfonts,amstext,amssymb,amsbsy,amsopn,amsthm,eucal}
\usepackage{latexsym,amsfonts,amssymb,amsmath,amsbsy}
\usepackage{graphics}

\newtheorem{Theorem}{Theorem}[section]

\newtheorem{Corollary}[Theorem]{Corollary}
\newtheorem{Proposition}[Theorem]{Proposition}

\newtheorem{exmp}[Theorem]{Example}

\newtheorem{rem}[Theorem]{Remark}
\newenvironment{Remark}{\begin{rem}\rm}{\end{rem}}

\newcommand{\dist}{\mathrm{dist}}

\newcommand{\spread}{\mathrm{spread}}
\newcommand{\Tr}{\mathrm{Tr}}
\newcommand{\triple}{\||}

 \DeclareMathOperator{\diag}{diag}
\date{}

\title{Modal approximations to damped linear systems
}

\author{Kre\v{s}imir Veseli\'{c}\thanks{Fakult\"{a}t f\"{u}r
Mathematik und Informatik, Fernuniversit\"{a}t Hagen, Postf 940, 58084 Hagen, Germany}
}

\begin{document}

\maketitle

\begin{abstract}We consider a finite dimensional damped second order system
and obtain spectral inclusion theorems for the related quadratic eigenvalue
problem. The inclusion sets are the 'quasi Cassini ovals'
which may greatly outperform standard Gershgorin circles. As the unperturbed
system we take a modally damped part of the system; this includes
the known proportionally damped models, but may give much sharper
estimates. These inclusions are then
applied to derive some easily calculable sufficient conditions
for the overdampedness of a given damped system.
\end{abstract}

\section{Introduction and preliminaries} A damped linear system without
gyroscopic forces is governed by the differential equation
\begin{equation}\label{1-MCK}
M \ddot x + C \dot x + Kx = f(t).
\end{equation}
Here $x=x(t)$ is an $\mathbb{R}^n$-valued  function of time
$t\in \mathbb{R}$; $M, C, K$ are real symmetric matrices of order
$n$. Typically $M, K$ are positive definite whereas $C$ is
positive semidefinite. The
physical meaning of these objects is
\begin{eqnarray*}
x(t) & \text{position or displacement}\\
M & \text{mass }\\
C & \text{damping }\\
K & \text{stiffness }\\
f(t)& \text{external force }
\end{eqnarray*}
If in the homogeneous equation above
we
insert $x(t)=e^{\lambda t}x$, $x$ constant, we obtain
\begin{equation}\label{qep}
(\lambda^2M+\lambda C+K)x=0
\end{equation}
which is called the quadratic eigenvalue problem,
attached to (\ref{1-MCK}), $\lambda$ is an eigenvalue and $x$
a corresponding eigenvector.

The quadratic eigenvalue problem may have poor spectral theory
in spite of the hermiticity and positive (semi)definiteness
of \(M.C,K\). There always exists a non-singular matrix
\(\Phi\) such that
\begin{equation}\label{modal_transf}
\Phi^TM\Phi = I,\quad
\Phi^TK\Phi = \Omega =
\diag(\omega_1^2,\ldots,\omega_n^2).
\end{equation}
If the matrix \(\Phi\) can be chosen such that also
\begin{equation}\label{modal_C}
D = \Phi^TC\Phi   
\end{equation}
is diagonal then the system is called {\em modally damped}.

While (\ref{modal_transf}) is the standard spectral decomposition
of a symmetric positive definite matrix pair.
a simultaneous achieving of (\ref{modal_C}) is
rather an exception being equivalent to the
generalised commutativity property
\begin{equation}\label{eq_modal}
CK^{-1}M=MK^{-1}C.
\end{equation}
However, as an approximation, modal damping is attractive
since it is handled by the standard theory and numerics of
Hermitian matrices. The aim of this paper is to assess
modal approximations of general damped systems. More precisely,
we will derive spectral inclusion theorems for eigenvalues
where the unperturbed system is modally damped. There is some
hierarchy among various modal approximations of a given
damped system and we will investigate this issue as well.
Our inclusion sets will not be circles, we will call them
quasi Cassini ovals. We will show that our ovals outdo
classical Gershgorin circles.
A special case are overdamped systems the eigenvalues of
which are particularly well behaved, there ovals reduce to
intervals and inclusions of Wielandt-Hoffman type will
be derived. Finally, we will derive new calculable sufficient
conditions for the overdampedness of a given system.

\section{Modal approximation}
  \label{Modal approximation}
Some, rather rough, facts on the positioning of
the eigenvalues are given
in \cite{l}. Further, more detailed, information
is obtained by the perturbation theory.
A simplest thoroughly known system is the undamped one.
Next to this lie the modally damped systems.

A simplest eigenvalue inclusion
for a general matrix \(A\) close to a matrix \(A_0\)
is
\begin{equation}\label{block_subsets}
\sigma(A) \subseteq {\cal G}_1 =
\{\lambda:\ \|(A - A_0)(A_0 - \lambda I)^{-1}\| < 1\}
\end{equation}
Obviously \({\cal G}_1 \subseteq {\cal G}_2\) with
\begin{equation}\label{Gershg2}
{\cal G}_2 =
\{\lambda:\ \|(A_0 - \lambda I)^{-1}\|^{-1} \leq \|(A - A_0)\|\}.
\end{equation}

This is valid for any matrices \(A,A_0\). Using \(\Phi\), \(\Omega\)
from (\ref{modal_transf}) we set
\[
y_1 = \Omega\Phi^{-T}x,\quad y_2 = \lambda\Phi^{-T}x,
\]
so the quadratic eigenvalue equation (\ref{qep}) is equivalent to
\begin{equation}\label{linep}
Ay = \lambda y.
\end{equation}
Here we have set
\begin{equation}\label{A_A0}
A  =
\left[\begin{array}{rr}
0        &   \Omega\\
-\Omega  &  -D\\
\end{array}\right],\quad
A_0 =
\left[\begin{array}{rr}
0       &  \Omega\\
-\Omega &    0\\
\end{array}\right].
\end{equation}
Hence
\[
A - A_0 =
\left[\begin{array}{rr}
0  &   0\\
0  &  -D\\
\end{array}\right].
\]
The matrix \(A_0\) is skew-symmetric and therefore normal, so
\(\|(A_0 - \lambda I)^{-1}\|^{-1} = \dist(\lambda,\sigma(A_0))\)
hence
\begin{equation}\label{Gershg2_norm}
{\cal G}_2 =
\{\lambda:\ \dist(\lambda,\sigma(A_0) \leq \|(A - A_0)\|\}
\end{equation}
where
\begin{equation}\label{maxCM}
\|(A - A_0)\| = \|D\|= \|L_2^{-1}CL_2^{-T}\| =
\max\frac{x^TCx}{x^TMx}
\end{equation}
is the largest eigenvalue of the
matrix pair \(C,M\). We may say that here 'the size
of the damping is measured relative to the mass'.

Thus, the perturbed eigenvalues are contained in the union of the
disks of radius \(\|D\|\) around \(\sigma(A_0)\).
\begin{Remark}\label{Rjundamped} In fact, \(\sigma(A)\)
is also contained in the union of the disks
\begin{equation}\label{Gershg_Rj}
\{\lambda:\ |\lambda \mp i\omega_j| \leq R_j\}
\end{equation}
with
\begin{equation}\label{Rj}
R_j = \sum_{k=1}^n|d_{kj}|.
\end{equation}
(Replace the spectral norm in (\ref{block_subsets}) by
the norm \(\|\cdot\|_1\)).
\end{Remark}
The bounds obtained above are, in fact, too crude,
since we have not taken into account the structure
of the perturbation \(A - A_0\) which has a remarkable
zero pattern.

Instead of working with the matrix \(A\) we may turn back to the
original quadratic eigenvalue problem in the representation in
the form
(see (\ref{modal_transf}) and (\ref{A_A0}))
\[
\det(\lambda^2 I + \lambda D + \Omega^2) = 0.
\]
The inverse
\[
(\lambda^2 I + \lambda D + \Omega^2)^{-1} =
\]
\[
(\lambda^2 I + \Omega^2)^{-1}
(I + \lambda D(\lambda^2 I + \Omega^2)^{-1})^{-1}
\]
exists, if
\begin{equation}\label{neum_1}
\|D(\lambda^2 I + \Omega^2)^{-1}\||\lambda|
< 1
\end{equation}
which is implied by
\begin{equation}\label{Dbounds}
\|(\lambda^2 I + \Omega^2)^{-1}\|\|D\||\lambda| =
\frac{\|D\||\lambda|}{\min_j(|\lambda - i\omega_j||\lambda + i\omega_j|)}
< 1
\end{equation}
Thus,
\begin{equation}\label{Dcassini}
\sigma(A) \subseteq \cup_j{\cal C}
(i\omega_j,-i\omega_j,\|D\|),
\end{equation}
where the set
\begin{equation}\label{C_ovals}
{\cal C}(\lambda_+,\lambda_-,r) =
\{\lambda: |\lambda - \lambda_+||\lambda - \lambda_-| \leq |\lambda|r\}
\end{equation}
will be called {\em quasi Cassini ovals} with foci \(\lambda_\pm\)
and extension \(r\). This is in analogy with the standard Cassini
ovals where on the right hand side
instead of \(|\lambda|r\) one has just \(r^2\). (The latter
also appear in eigenvalue bounds in somewhat different context.)
We note the obvious relation
\begin{equation}\label{C_ov_ord}
{\cal C}(\lambda_+,\lambda_-,r) \subset
{\cal C}(\lambda_+,\lambda_-,r'),\quad \mbox{whenever }
r < r'.
\end{equation}
The
quasi Cassini ovals are qualitatively similar to the
standard ones; they can consist of one or two components;
the latter case occurs when \(r\) is sufficiently small
with respect to \(|\lambda_+ - \lambda_-|\). In this case
the ovals in (\ref{Dcassini}) are approximated by the
disks
\begin{equation}\label{Dhalf}
|\lambda \pm i\omega_j| \leq \frac{\|D\|}{2}
\end{equation}
and this is one half of the bound in
(\ref{Gershg2_norm}), (\ref{maxCM}).
\begin{Remark}\label{rel_ovals}  \(\sigma(A)\)
is also contained in the union of the ovals
\begin{equation}\label{eq:rel_ovals}
{\cal C}(i\omega_j,-i\omega_j,\|\Omega^{-1}D\Omega^{-1}\|\omega_j^2).
\end{equation}
Indeed, instead of inverting \(\lambda^2 I + \lambda D + \Omega^2\)
invert \(\lambda^2  \Omega^{-2} + \lambda \Omega^{-1}D\Omega^{-1} + I\).
\end{Remark}
\begin{Remark}\label{Rjund_ovals}  \(\sigma(A)\)
is also contained in the union of the ovals
\begin{equation}\label{Gershg_Rj_ovals}
{\cal C}(i\omega_j,-i\omega_j,R_j)
\end{equation}
and also
\begin{equation}\label{rel_Rj_ovals}
{\cal C}(i\omega_j,-i\omega_j,\rho_j\omega_j^2)
\end{equation}
with
\begin{equation}\label{rhoj}
\rho_j = \sum_{{k=1 \atop k\neq j}}^n\frac{|d_{kj}|}{\omega_k\omega_j}.
\end{equation}
\end{Remark}

The just considered undamped approximation was just a prelude
to the main topic of this section, namely the modal approximation.
The modally damped systems are so much simpler than the general ones that
practitioners often substitute the true damping matrix
by some kind of 'modal approximation'. Most typical such approximations
in use are of the form
\begin{equation}\label{C_proport}
C_{prop} = \alpha M +  \beta K
\end{equation}
where \(\alpha,\beta\) are chosen in such a way that \(C_{prop}\)
be in some sense as close as possible to \(C\), for instance,
\begin{equation}\label{C_proport_E}
\Tr\left[(C - \alpha M - \beta K)W(C - \alpha M - \beta K)\right]
= \min,
\end{equation}
where \(W\) is some convenient positive definite weight matrix.
This is a {\em proportional approximation}.
In general such approximations may go quite astray and yield
thoroughly false predictions. We will now assess
them in a more systematic way.\\

A modal approximation to the system
(\ref{1-MCK}) is obtained by first representing it in modal coordinates
by the matrices \(D\), \(\Omega\) and then by replacing \(D\)
by its diagonal part
\begin{equation}\label{D0}
D^0 = \diag(d_{11},\ldots,d_{nn}).
\end{equation}
The off-diagonal part \(D' = D - D^0\) is considered a perturbation.
Again we can work in the phase space or with the original
quadratic eigenvalue formulation. In the first case
we can make perfect shuffling to obtain
\begin{equation}\label{A_shuffled}
A = (A_{i,j}),\quad
A_{ii} =
\left[\begin{array}{rr}
0         &  \omega_i \\
-\omega_i & d_{ii}    \\
\end{array}\right],\quad
A_{ij} =
\left[\begin{array}{rr}
0     &  0     \\
0     & d_{ij} \\
\end{array}\right]
\end{equation}
\begin{equation}\label{A0_shuff_mod}
A_0 = \diag(A_{11},\ldots,A_{nn}).
\end{equation}
So, for \(n = 3\)
\[
A =
\left[\begin{array}{cc|cc|cc}
0        &\omega_1 & 0           & 0       & 0         &       0\\
-\omega_1& -d_{11} & 0           & -d_{12} & 0         & -d_{13}\\\hline
0        &   0     & 0           &\omega_2 & 0         & 0      \\
0        & -d_{12} & -\omega_2   & -d_{22} & 0         & -d_{23}\\\hline
0        &  0      &   0         & 0       & 0         &\omega_3\\
0        & -d_{13} & 0           & -d_{23} & -\omega_3 & -d_{33}\\
\end{array}\right].
\]
 Then
\[
\|(A_0 - \lambda I)^{-1}\|^{-1} =
\max_j\|(A_{jj} - \lambda I)^{-1}\|^{-1}.
\]
Even for \(2\times 2\)-blocks any common norm of
\((A_{jj} - \lambda I)^{-1}\) seems complicated
to express in terms of disks or other simple regions,
unless we diagonalise each \(A_{jj}\) as
\begin{equation}\label{Sjlambda}
S_j^{-1}A_{jj}S_j =
\left[\begin{array}{rr}
\lambda_+^j    &  0     \\
0     & \lambda_-^j\\
\end{array}\right],\quad
\lambda_\pm^j = \frac{-d_{jj} \pm \sqrt{d_{jj}^2 - 4\omega_j^2}}{2}.
\end{equation}
As is directly verified,
\[
\kappa(S_j) = \sqrt{\frac{1 + \theta_j^2}{|1 - \theta_j^2|}},\quad
\theta_j = \frac{d_{jj}}{2\omega_j}.
\]
with
\[
\theta_j = \frac{d_{jj}}{2\omega_j}.
\]
Set \(S = \diag(S_{11},\ldots,S_{nn})\) and
\[
A' = S^{-1}AS = A_0' + A''
\]
then
\[
A_0' = \diag(\lambda_\pm^1,\ldots,\lambda_\pm^n),
\]
\[
A_{jk}'' = S_j^{-1}A_{jk}'S_k,\quad A'' = S^{-1}A'S
\]
Now the general perturbation bound (\ref{Gershg2_norm}), applied
to \(A_0', A''\), gives
\begin{equation}\label{S_bound}
\sigma(A) \subseteq
\cup_{j,\pm}\{\lambda:\ |\lambda - \lambda_\pm^j| \leq \kappa(S)\|D'\|\}.
\end{equation}

There is a related 'Gershgorin-type bound'
\begin{equation}\label{Gershg_Rj_Sj}
\sigma(A) \subseteq
\cup_{j,\pm}\{\lambda:\ |\lambda - \lambda_\pm^j| \leq \kappa(S_j)r_j\}
\end{equation}
with
\begin{equation}\label{r_j}
r_j = \sum_{k=1\atop j\neq i}^n \|d_{jk}\|.
\end{equation}
To show this we replace the spectral norm \(\|\cdot\|\)
in (\ref{block_subsets}) by the norm \(\triple \cdot \triple_1\),
defined as
\[
\triple A \triple_1: = \max_j\sum_k\|A_{kj}\|
\]
where the norms on the right hand side are spectral.
Thus, (\ref{block_subsets}) will hold, if
\[
\max_j\sum_k\|(A - A_0)_{kj}\|\|(A_{jj} - \lambda I)^{-1}\|
< 1
\]
Taking into account the equality
\[
\|(A - A_0)_{kj}\| =
\left\{
\begin{array}{rr}
|d_{kj}|,  &   k \neq j\\
0          &   k = j
\end{array}\right.
\]
\(\lambda \in\sigma(A)\) implies
\[
r_j \geq \|(A_{jj} - \lambda I)^{-1}\| \geq
\frac{\min\{|\lambda - \lambda_+^j|,|\lambda - \lambda_-^j|\}}
{\kappa(S_j)}
\]
and this is (\ref{Gershg_Rj_Sj}).\\

Note that the bounds (\ref{S_bound}) and (\ref{Gershg_Rj_Sj})
are poor whenever the modal approximation
is close to a critically damped eigenvalue.\\

Better bounds are expected, if we work
directly with the quadratic eigenvalue equation.
The inverse
\[
(\lambda^2 I + \lambda D + \Omega^2)^{-1} =
\]
\[
(\lambda^2 I + \lambda D^0 + \Omega^2)^{-1}
(I + \lambda D'(\lambda^2 I + \lambda D^0 + \Omega^2)^{-1})^{-1}
\]
exists, if
\begin{equation}\label{neum_2}
\|D'(\lambda^2 I + \lambda D^0 + \Omega^2)^{-1}\||\lambda|
< 1
\end{equation}
which is insured, if
\begin{equation}\label{D'bounds}
\|(\lambda^2 I  + \lambda D^0 + \Omega^2)^{-1}\|\|D'\||\lambda| =
\frac{\|D'\||\lambda|}{\min_j(|\lambda - \lambda_+^j||\lambda - \lambda_-^j|)}
< 1
\end{equation}
Thus,
\begin{equation}\label{D'cassini}
\sigma(A) \subseteq \cup_j{\cal C}
(\lambda_+^j,\lambda_-^j,\|D'\|).
\end{equation}
These ovals will always have both foci either real
or complex conjugate. If \(r = \|D'\|\)
is small with respect to
\(|\lambda_+^j - \lambda_-^j| = \sqrt{|d_{jj}^2 - 4\omega_j^2|}\)
then either \(|\lambda - \lambda_+^j|\) or \(|\lambda - \lambda_-^j|\)
is small. In the first case the inequality
\(|\lambda - \lambda_+^j||\lambda - \lambda_-^j| \leq
|\lambda|r\) is approximated by
\begin{equation}\label{C_ovals_appr+}
|\lambda - \lambda_+^j|
\leq \frac{|\lambda_+^j|r}{|\lambda_+^j - \lambda_-^j|} =
r\left\{\begin{array}{ll}
\frac{\omega_j}{\sqrt{d_{jj}^2 - 4\omega_j^2}} & d_{jj} < 2\omega_j\\
\mbox{}\\
\frac{d_{jj} - \sqrt{d_{jj}^2 - 4\omega_j^2}}
{\sqrt{d_{jj}^2 - 4\omega_j^2}} & d_{jj} > 2\omega_j\\
\end{array}\right.
\end{equation}
and in the second
\begin{equation}\label{C_ovals_appr-}
|\lambda - \lambda_-^j|
\leq \frac{|\lambda_-^j|r}{|\lambda_+^j - \lambda_-^j|} =
r\left\{\begin{array}{ll}
\frac{\omega_j}{\sqrt{d_{jj}^2 - 4\omega_j^2}} & d_{jj} < 2\omega_j\\
\mbox{}\\
\frac{d_{jj} + \sqrt{d_{jj}^2 - 4\omega_j^2}}
{\sqrt{d_{jj}^2 - 4\omega_j^2}} & d_{jj} > 2\omega_j\\
\end{array}\right. .
\end{equation}
This is again a union of  disks. If \(d_{jj} \approx 0\)
then their radius is \(\approx r/2\).
If \(d_{jj} \approx 2\omega_j\)
i.e.~\(\lambda_- = \lambda_+ \approx -d_{jj}/2\)
the ovals look like a single
circular disk. For large \(d_{jj}\) the oval around the absolutely
larger eigenvalue is \(\approx r\) (the same
behaviour as with (\ref{Gershg_Rj_Sj})) whereas the
smaller eigenvalue has the diameter
\(\approx 2r\omega_j^2/d_{jj}^2\) which is drastically better than
(\ref{Gershg_Rj_Sj}).

In the same way as before the Gershgorin type estimate
is obtained
\begin{equation}\label{rj_cassini}
\sigma(A) \subseteq \cup_j{\cal C}
(\lambda_+^j,\lambda_-^j,r_j).
\end{equation}

We have called \(D'\) {\em a} modal approximation to
\(D\) because the matrix \(D\) is not uniquely determined
by the input matrices \(M,C,K\).
Different choices of the transformation matrix
\(\Phi\) give rise to different modal approximations
\(D'\) but the differences between them are mostly non-essential.
To be more precise, let
\(\Phi\) and \(\tilde{\Phi}\) both satisfy
(\ref{modal_transf}). Then
\[
M = \Phi^{-T}\Phi^{-1} = \tilde{\Phi}^{-T}\tilde{\Phi}^{-1},
\]
\[
K = \Phi^{-T}\Omega^2\Phi^{-1} = \tilde{\Phi}^{-T}\Omega^2\tilde{\Phi}^{-1}
\]
implies that \(U = \Phi^{-1}\tilde{\Phi}\) is an
orthogonal matrix which commutes
with
\begin{equation}\label{Omega_mult}
\Omega = \diag(\omega_1I_{n_1},\ldots,\omega_sI_{n_s}),
\quad \omega_1 < \cdots < \omega_s.
\end{equation}
Hence
\[
U^0 = \diag(U_{11},\ldots,U_{ss}),
\]
where each \(U_{jj}\) is an orthogonal matrix of order \(n_j\) from (\ref{D0}). Now,
\begin{equation}\label{D_similarity}
\tilde{D} = \tilde{\Phi}^T C \tilde{\Phi} = U^T \Phi^T C \Phi U = U^T D U,
\end{equation}
\begin{equation}\label{Dij_similarity}
\tilde{D}_{ij} = U_i^TD_{ij}U_j
\end{equation}
and hence
\begin{equation}\label{D'_similarity}
\tilde{D}' = U^TD'U.
\end{equation}
Now, if the undamped frequencies are all simple,
then \(U\) is diagonal and
the estimates (\ref{maxCM}) or  (\ref{Dcassini})--(\ref{C_ovals})
remain unaffected by this change of coordinates. Otherwise
we replace \(\diag(d_{11},\ldots,d_{nn})\) by
\begin{equation}\label{DD0}
D^0 = \diag(D_{11},\ldots,D_{ss})
\end{equation}
where \(D^0\) commutes with \(\Omega\).
In fact, a general definition of a modal approximation
is that it
\begin{enumerate}
\item is block-diagonal and
\item commutes with
\(\Omega\).
\end{enumerate}
The modal approximation with the coarsest
possible partition --- this is the one whose block dimensions
equal the multiplicities in \(\Omega\) --- is
called a {\em maximal modal approximation}.
Accordingly, we say that
\(C^0 =\Phi^{-1}D^0\Phi^{-T}\) is a modal approximation to
 \(C\) (and also \(M,C^0,K\) to
\(M,C,K\)).
\begin{Proposition}\label{all_modals}
Each modal approximation to \(C\)
is of the form
\begin{equation}\label{modal_form}
C^0 = \sum_{k=1}^sP_k^*CP_k
\end{equation}
where \(P_1,\ldots.P_s\) is an \(M\)-orthogonal decomposition
of the identity (that is \(P_k^* = MP_kM^{-1}\)) and
\(P_k\) commute with the matrix
\[
\sqrt{M^{-1}K} = M^{-1/2}\sqrt{M^{-1/2}KM^{-1/2}}M^{1/2}
\]
\end{Proposition}
{\bf Proof.} Use the formula
\[
D^0 = \sum_{k=1}^sP_k^0DP_k^0
\]
with
\[
P_k^0 = \diag(0\ldots,I_{n_k},\ldots,0,\quad
D = \Phi^TC\Phi,\quad D^0 = \Phi^TC^0\Phi
\]
and set \(P_k = \Phi P_k^0\Phi^{-1}\). Q.E.D.

It is obvious that the maximal approximation is the best among
all modal approximations in the sense that
\begin{equation}\label{modal_E_best}
\|D - D^0\|_E \leq \|D - \hat{D}^{0}\|_E,
\end{equation}
where
\begin{equation}\label{D_fine}
\hat{D}^{0} = \diag(\hat{D}_{11},\ldots,\hat{D}_{zz})
\end{equation}
and \(D = (\hat{D}_{ij})\) is any block partition of \(D\) which
is finer than that in (\ref{DD0}).
We will now prove that the inequality
(\ref{modal_E_best}) is valid for the spectral norm also. We
shall need the following
\begin{Proposition}\label{prop_offd_estimate}
Let \(H = (H_{ij})\) be any partitioned Hermitian matrix such
that the diagonal blocks \(H_{ii}\) are square. Set
\[
H^0 = \diag(H_{11},\ldots,H_{ss}),\quad
H' = H - H^0.
\]
Then
\begin{equation}\label{offd_estimate}
\lambda_k(H) - \lambda_n(H) \leq \lambda_k(H')
\leq \lambda_k(H) - \lambda_1(H)
\end{equation}
where \(\lambda_k(\cdot)\) denotes the non-decreasing sequence of the eigenvalues of any
Hermitian matrix.
\end{Proposition}
{\bf Proof.} By the monotonicity property (Wielandt's theorem) we have
\[
\lambda_k(H) - \max_j\max\sigma(H_{jj}) \leq \lambda_k(H')
\leq \lambda_k(H) - \min_j\min\sigma(H_{jj}).
\]
By the interlacing property,
\[
\lambda_1(H) \leq \sigma(H_{jj}) \leq \lambda_n(H).
\]
Together we obtain (\ref{offd_estimate}). Q.E.D.\\

From (\ref{offd_estimate}) some simpler estimates immediately follow:
\begin{equation}\label{offd_spread}
\|H'\|
\leq \lambda_n(H) - \lambda_1(H) =: \spread(H)
\end{equation}
and, if \(H\) is positive (or negative) semidefinite
\begin{equation}\label{offd_norm}
\|H'\|
\leq \|H\|.
\end{equation}

Now (\ref{modal_E_best}) for the spectral norm immediately follows from
(\ref{offd_norm}). So, a best bound in (\ref{D'cassini})
is obtained, if \(D^0 = D - D'\) is a maximal modal approximation.
\begin{Proposition}\label{proport_modal}
Any modal approximation is better than any proportional one.
\end{Proposition}
{\bf Proof.} With
\[
D_{\mbox{{\it\small prop}}} = \alpha I + \beta \Omega
\]
we have
\[
|(D - D_{\mbox{{\it \small prop}}})_{ij}| \geq
|(D - D^0)_{ij}| = |D^0_{ij}|
\]
which implies
\[
\|D - D_{\mbox{{\it\small  prop}}}\| \geq \|D'\|.
\]
Q.E.D.\\

If \(D^0\) is block diagonal and
the
corresponding \(D' = D - D^0\) is inserted in (\ref{D'cassini})
then the values \(d_{jj}\) from (\ref{Sjlambda}) should be replaced
by the corresponding eigenvalues of the diagonal blocks
\(D_{jj}\). But in this case we can further transform
\(\Omega\) and \(D\) by a unitary similarity
\[
U = \diag(U_1,\ldots,U_s)
\]
such that each of the blocks \(D_{jj}\) becomes diagonal
(\(\Omega\) stays unchanged). With this stipulation
we may retain the formula (\ref{D'cassini}) unaltered.
This shows that taking just the diagonal part \(D^0\) of
\(D\) covers, in fact, {\em all possible modal approximations},
when \(\Phi\) varies over all matrices performing
(\ref{modal_transf}).

Similar extension can be made with the bound (\ref{rj_cassini})
but then no improvements in general can be guaranteed although
they are more likely than not.\\

By the usual continuity argument it is seen that
the number of the eigenvalues in each component
of \(\cup_i{\cal C}(\lambda_+^i,\lambda_-^i,r_i)\)
is twice the number of involved diagonals. In particular,
if we have the maximal number of \(2n\) components,
then each of them contains
exactly one eigenvalue.\\
%

A strengthening in the sense of Brauer is possible as well.
We will show that the {\em spectrum is contained in the union of
double ovals}, defined as%
\[
{\cal D}(\lambda_+^{p},\lambda_-^{p},
\lambda_+^{q},\lambda_-^{q},r_pr_q)
=
\]
\begin{equation}\label{double_ovals}
\{\lambda: |\lambda - \lambda_+^{p}||\lambda - \lambda_-^{p}||\lambda -
\lambda_+^{q}||\lambda - \lambda_-^{q}| \leq r_pr_q|\lambda|^2\},
\end{equation}
where the union is taken over all pairs
\(p \neq q\) and \(\lambda_\pm^{p}\)
are the solutions of \(\lambda^2 + d_{pp}\lambda + \omega_p^2 = 0\)
and similarly for \(\lambda_\pm^{q}\). The proof just mimics
the standard Brauer's one. The quadratic eigenvalue problem is written
as
\begin{equation}\label{D'x_components_p}
(\lambda^2 + \lambda d_{pp} + \omega_i^2)x_i =
-\lambda\sum_{j=1\atop J\neq i}^n d_{ij}x_j,
\end{equation}
\begin{equation}\label{D'x_components_q}
(\lambda^2 + \lambda d_{ii} + \omega_i^2)x_i =
-\lambda\sum_{j=1\atop J\neq i}^n d_{ij}x_j,
\end{equation}
where \(|x_p| \geq |x_q|\) are the two absolutely largest components
of \(x\). If \(x_q = 0\) then \(x_j = 0\) for all \(j \neq p\) and trivially
\(\lambda \in
{\cal D}(\lambda_+^{p},\lambda_-^{p},\lambda_+^{q},\lambda_-^{q},r_pr_q) \).
If \(x_q \not= 0\) then multiplying the equalities
(\ref{D'x_components_p}) and (\ref{D'x_components_q}) yields
\[
|\lambda - \lambda_+^{p}||\lambda - \lambda_-^{p}||\lambda -
\lambda_+^{q}||\lambda - \lambda_-^{q}||x_p||x_q| \leq
\]
\[
|\lambda|^2\sum_{j=1\atop j\neq p}^n\sum_{k=1\atop k\neq q}^n
|d_{pj}||d_{qk}||x_j||x_k|.
\]
Because in the double sum above there is no term with
\(j = k = p\) we always have \(|x_j||x_k| \leq |x_p||x_q|\),
hence the said sum is bounded by
\[
|\lambda|^2|x_p||x_q|\sum_{j=1\atop j\neq p}^n|d_{pj}|
\sum_{k=1\atop k\neq q}^n
|d_{qk}|.
\]
Thus, our inclusion is proved. As it is immediately seen, {\em the union
of all double ovals is contained in the union of all quasi Cassini
ovals.}\\

The simplicity of the modal approximation suggests to try to extend
it to as many systems as possible. A close candidate for such extension
is any system with tightly clustered undamped frequencies, that is,
\(\Omega\) is close to an \(\Omega^0\) from (\ref{Omega_mult}).
Starting again with
\[
(\lambda^2 I + \lambda D + \Omega^2)^{-1} =
\]
\[
(\lambda^2 I + \lambda D^0 + (\Omega^0)^2)^{-1}
(I + (\lambda D' + Z)+
(\lambda^2 I + \lambda D^0 + (\Omega^0)^2)^{-1})^{-1}
\]
with \(Z = \Omega^2 - (\Omega^0)^2\) we immediately obtain
\begin{equation}\label{D'cassini_mod}
\sigma(A) \subseteq \cup_j\hat{{\cal C}}
(\lambda_+^j,\lambda_-^j,\|D'\|,\|Z\|).
\end{equation}
where the set
\begin{equation}\label{C_ovals_mod}
\hat{{\cal C}}(\lambda_+,\lambda_-,r,q) =
\{\lambda: |\lambda - \lambda_+||\lambda - \lambda_-| \leq |\lambda|r + q\}
\end{equation}
will be called {\em modified Cassini ovals} with foci \(\lambda_\pm\)
and extensions \(r,q\).
\begin{figure}[htp]
\begin{center}\centering
\includegraphics[width=3.5cm,height=9cm]{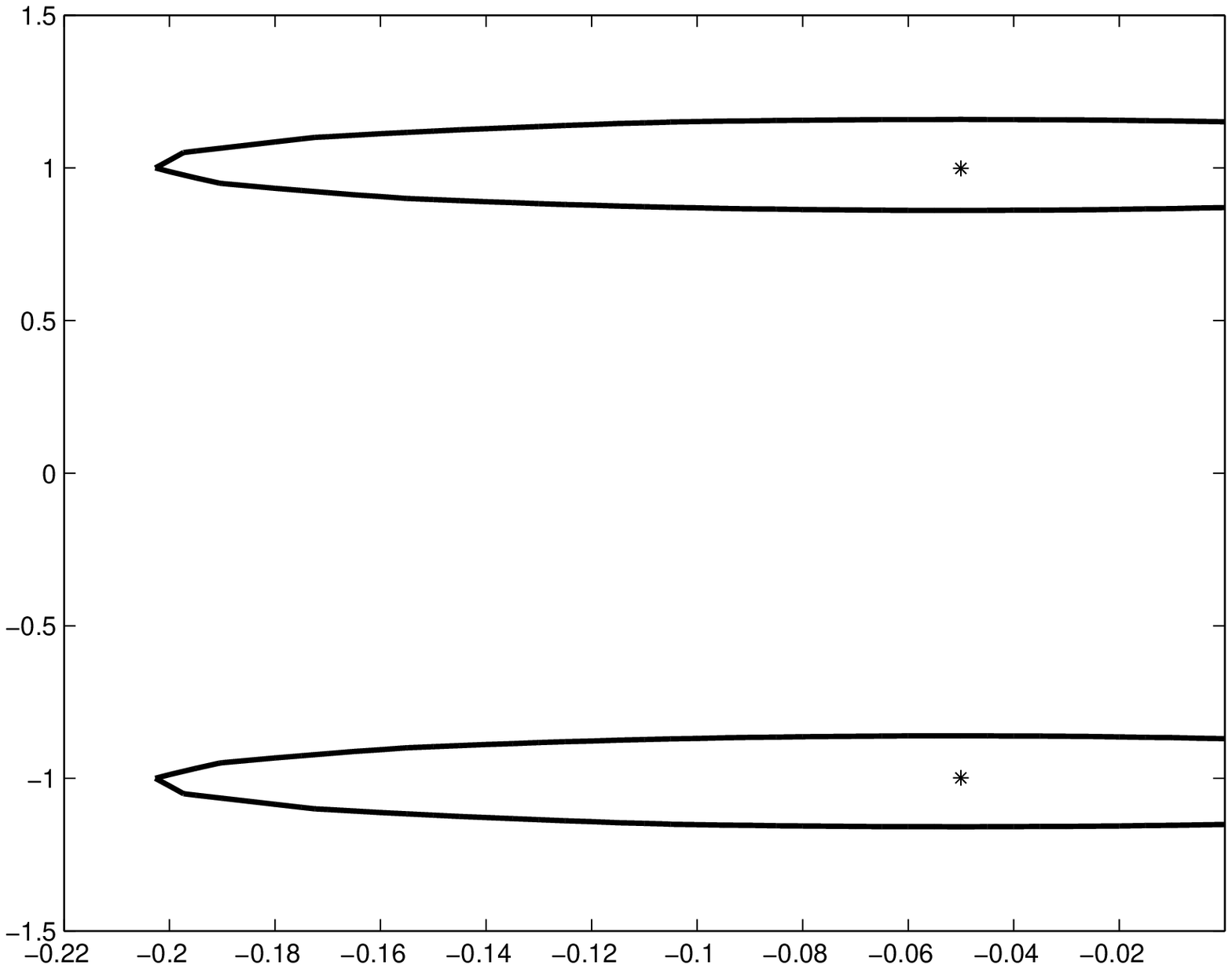}
\hspace{2cm}
\includegraphics[width=4.8cm,height=9cm]{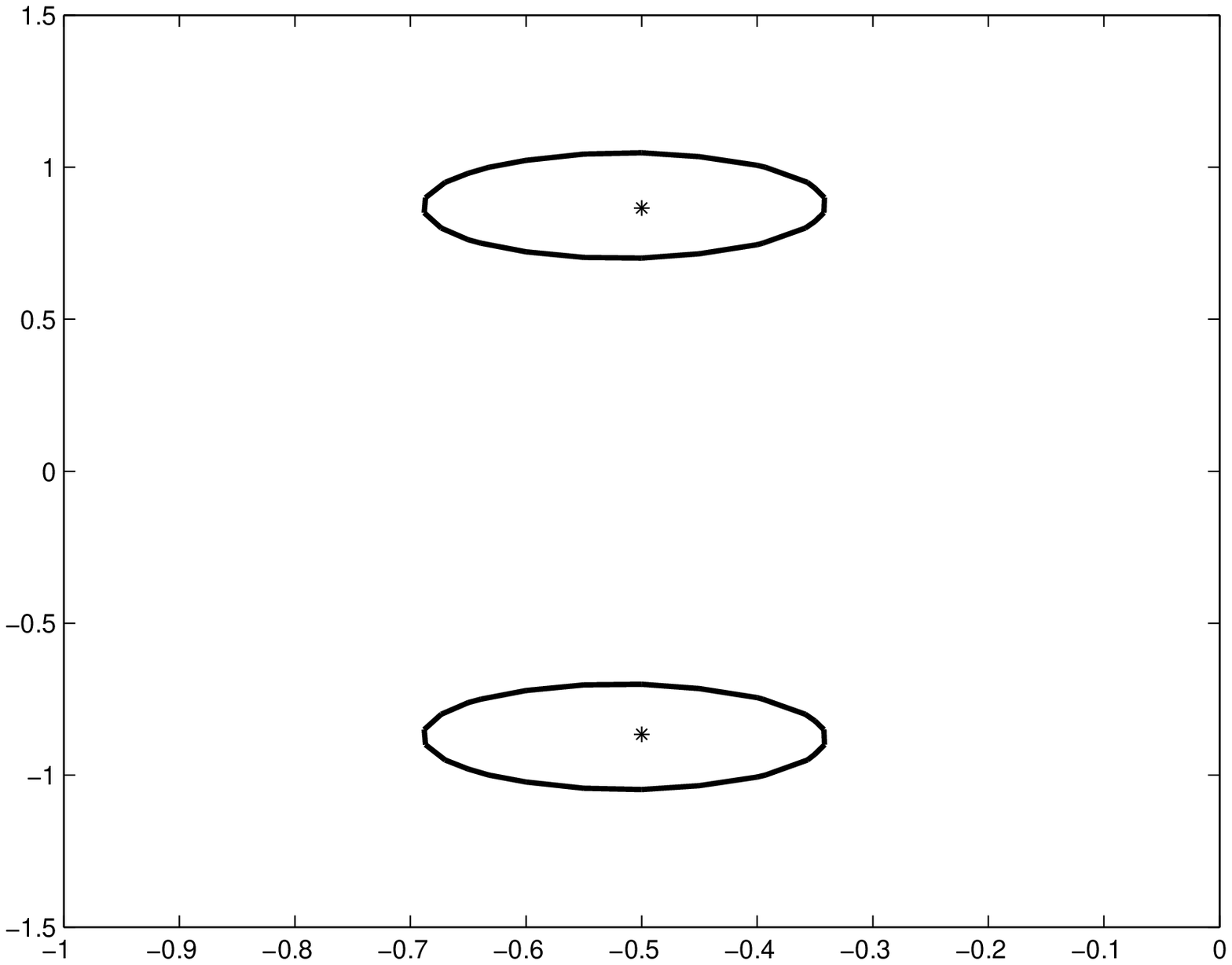}
\caption{Ovals for \(\omega=1;d=0.1,1;\,r=0.3\)}
 \label{disks_ovals12}
\end{center}
\end{figure}
\begin{figure}[htp]
\begin{center}\centering
\includegraphics[width=4cm,height=9cm]{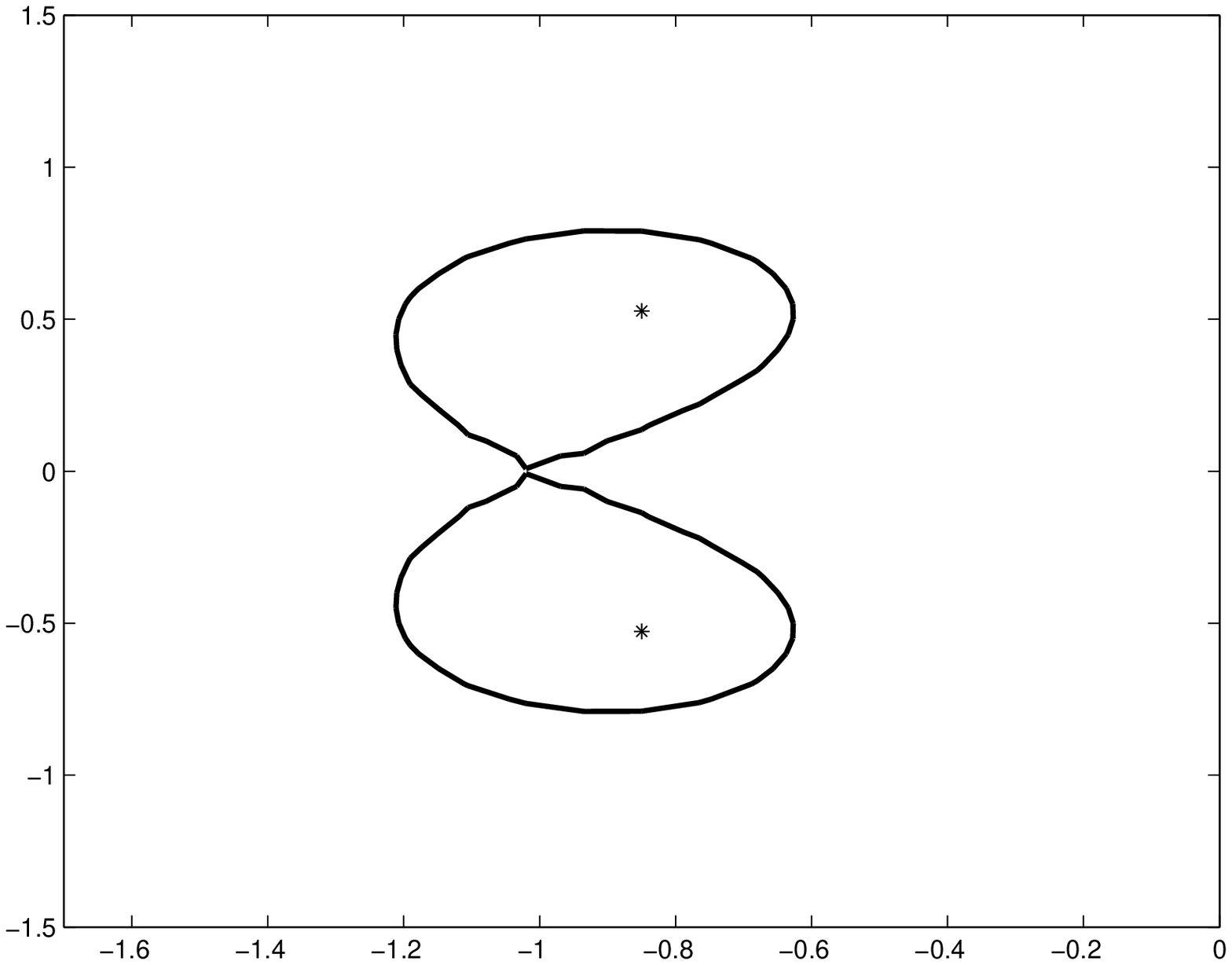}
\hspace{0.5cm}
\includegraphics[width=4cm,height=9cm]{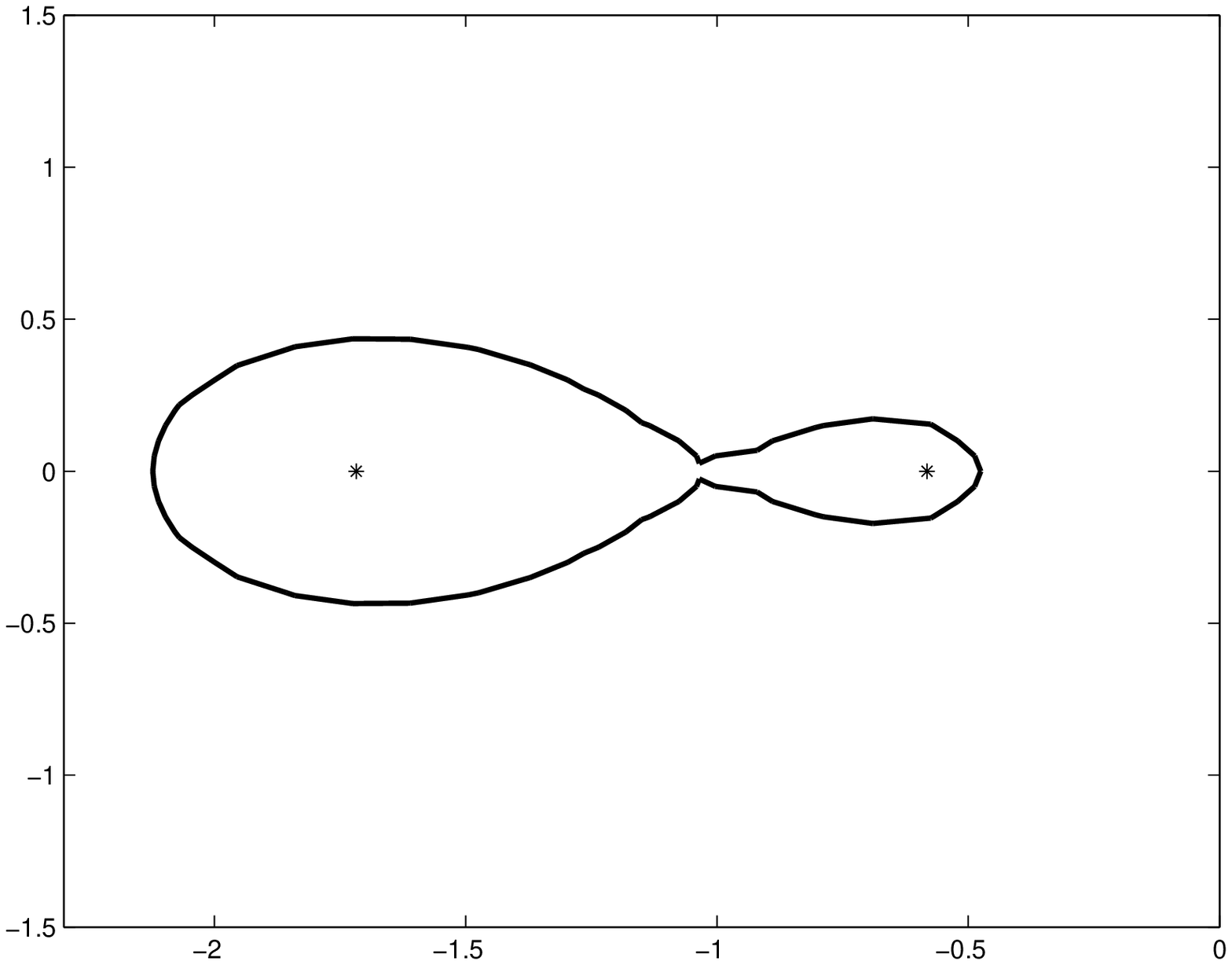}
\hspace{0.5cm}
\includegraphics[width=4cm,height=9cm]{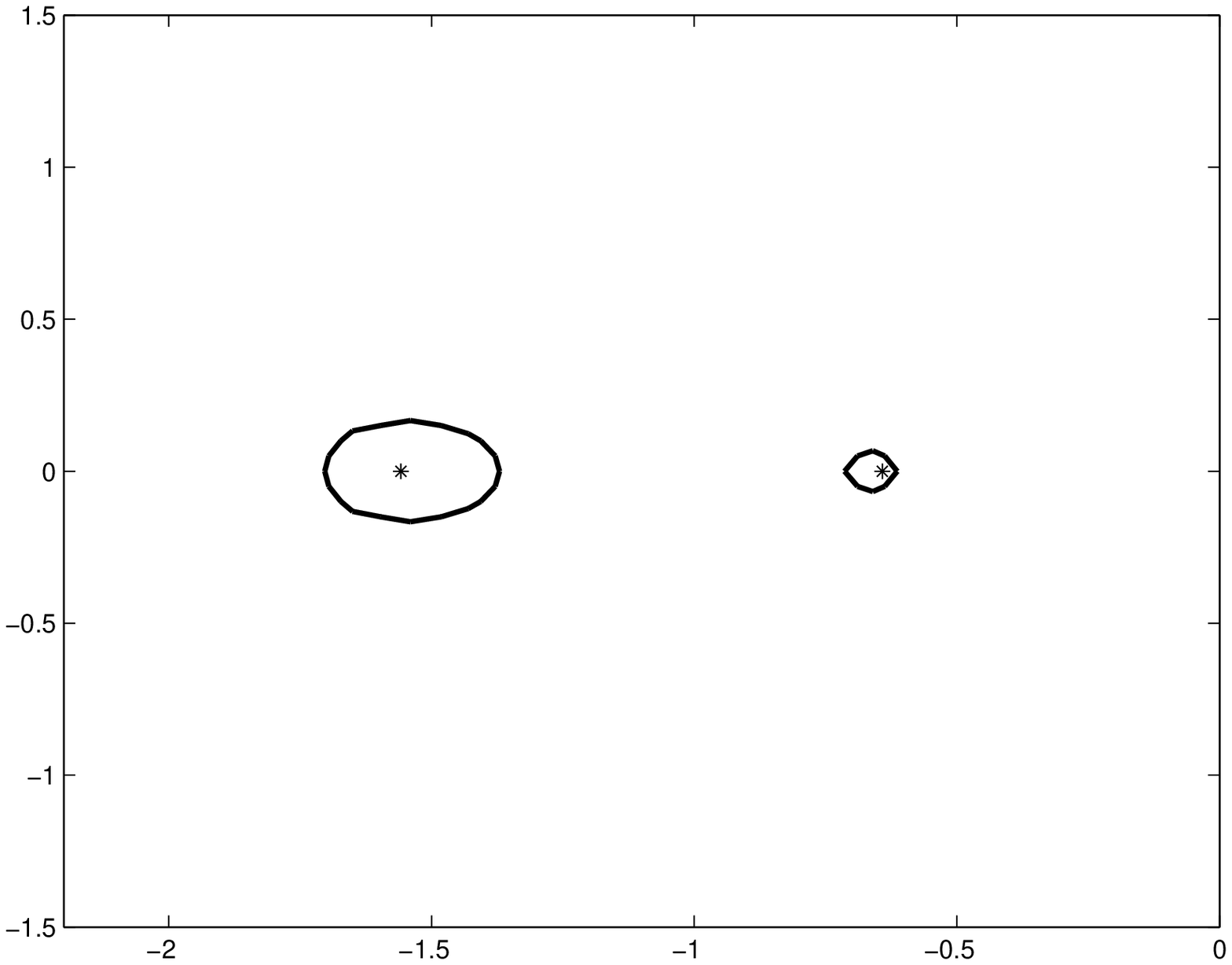}
\caption{Ovals for \(\omega=1;\,d=1.7,2.3,2.2;\,r=0.3,0.3,0.1\)}
 \label{disks_ovals345}
\end{center}
\end{figure}

\begin{Remark}\label{modal_circuits}
The basis of any modal approximation
is the diagonalisation of the matrix pair
\(M,K\). An analogous procedure
with similar results can be performed by diagonalising
the pair \(M,K\) or \(C,K\).
\end{Remark}

\section{Modal approximation and overdampedness}
  \label{Modal approximation and overdampedness}
If the systems in the previous section are all overdamped
then estimates are greatly simplified as ovals become
just intervals. But before going into this a more elementary
--- and more important --- question arises: Can the
modal approximation help to decide the overdampedness of a
given system?\\

We begin with some obvious facts the proofs
of which are left to the reader.

\begin{Proposition}\label{overd_ritz}If the system \(M,C,K\)
is overdamped, then the same is true of the
{\em projected system}
\begin{equation}\label{MCK_projected}
M' = X^*MX,\quad C' = X^*CX,\quad K' = X^*KX
\end{equation}
where \(X\) is any injective matrix. Moreover,
the definiteness interval of the former is
contained in the one of the latter.
\end{Proposition}
\begin{Proposition}\label{block_overd}
Let
\[
M = \diag(M_{11},\ldots,M_{ss})
\]
\[
C = \diag(C_{11},\ldots,C_{ss})
\]
\[
K = \diag(K_{11},\ldots,K_{ss}).
\]
Then the system \(M,C,K\) is overdamped, if and only if
each of \(M_{jj},C_{jj},K_{jj}\) is overdamped and
their definiteness intervals have a non trivial intersection
(which is then the definiteness interval of \(M,C,K\))
\end{Proposition}
\begin{Corollary}\label{modal_overd}If the system \(M,C,K\)
is overdamped, then the same is true of any of its
modal approximations.
\end{Corollary}
Obviously, if a maximal modal
approximation is overdamped, then so are all others.

In the following we shall need some well known
sufficient conditions for negative definiteness
of a general Hermitian matrix \(A = (a_{ij})\);
these are:
\begin{equation}\label{ajj<0}
a_{jj} < 0
\end{equation}
 for all \(j\) and either
\begin{equation}\label{norm_dominance}
\|A - \diag(a_{11},\ldots,a_{nn})\| <
       -\max_j a_{jj}
\end{equation}
 (norm-diagonal dominance) or
\begin{equation}\label{Gershg_dominance}
\sum_{k=1\atop k\neq j}^n|a_{kj}| <
       -a_{jj} \mbox{ for all } j
\end{equation}
(Gershgorin-diagonal dominance).
\begin{Theorem}\label{th_modal_overd}Let \(\Omega\), \(D\),
\(r_j\)
be from (\ref{modal_transf}), (\ref{modal_C}),  (\ref{r_j}), respectively
and
\[
D^0 = \diag(d_{11},\ldots,d_{nn}),\quad D' = D - D^0.
\]
Let either
\begin{equation}\label{Deltaj}
\Delta_j = (d_{jj} - \|D'\|)^2 - 4\omega_j^2 > 0
\mbox{ for all } j
\end{equation}
and
\begin{equation}\label{p-+}
p_- := \max_j\frac{-d_{jj} + \|D'\| - \sqrt{\Delta_j}}{2}
< \min_j\frac{-d_{jj} + \|D'\| + \sqrt{\Delta_j}}{2}
=: p_+
\end{equation}
or
\begin{equation}\label{hat_Deltaj}
\hat{\Delta}_j = (d_{jj} - r_j)^2 - 4\omega_j^2 > 0
\mbox{ for all } j
\end{equation}
and
\begin{equation}\label{hat_p-+}
\hat{p}_- := \max_j\frac{-d_{jj} + r_j - \sqrt{\hat{\Delta}_j}}{2}
< \min_j\frac{-d_{jj} + r_j + \sqrt{\hat{\Delta}_j}}{2}
=: \hat{p}_+.
\end{equation}
Then the system \(M,C,K\) is overdamped. Moreover, the interval
\((p_-, p_+)\), \((\hat{p}_-, \hat{p}_+)\), respectively,
is contained in the definiteness interval of \(M,C,K\).
\end{Theorem}
{\bf Proof.} Let \(p_- < \mu < p_+\). The negative definiteness
of
\[
\mu^2I + \mu D + \Omega^2 =
\mu^2I + \mu D^0 + \Omega^2 + \mu D'
\]
will be insured by norm-diagonal dominance, if
\[
-\mu\|D'\| < -\mu^2 - \mu d_{jj} - \omega_j^2
\mbox{ for all } j,
\]
that is, if \(\mu\) lies between the roots of the quadratic equation
\[
\mu^2 + \mu(d_{jj} - \|D'\|) + \omega_j^2 = 0
\mbox{ for all } j
\]
and this is insured by
(\ref{Deltaj}) and (\ref{p-+}). The conditions
(\ref{hat_Deltaj}) and (\ref{hat_p-+}) are treated analogously.
Q.E.D.\\

We are now prepared to adapt the spectral inclusion bounds
from the previous section to overdamped systems. Recall
that in this case the definiteness interval divides
the \(2n\) eigenvalues into two groups: \(J\)-negative
and \(J\)-positive.
\begin{Theorem}\label{th_overd_incl}If (\ref{Deltaj})
and (\ref{p-+}) hold then the \(J\)-negative/\(J\)-positive
eigenvalues are contained in
\begin{equation}\label{cupcup}
\cup_j(\mu_{--}^j,\mu_{-+}^j),\quad
\cup_j(\mu_{+-}^j,\mu_{++}^j),
\end{equation}
respectively, with
\begin{equation}\label{mu++--}
\mu_{++\atop --}^j =
\frac{-d_{jj} - \|D'\| \pm \sqrt{(d_{jj} + \|D'\|)^2 - 4\omega_j^2}}{2}
\end{equation}
\begin{equation}\label{mu+--+}
\mu_{+-\atop -+}^j =
\frac{-d_{jj} + \|D'\| \pm \sqrt{(d_{jj} - \|D'\|)^2 - 4\omega_j^2}}{2}.
\end{equation}
An analogous statement holds, if (\ref{hat_Deltaj})
and (\ref{hat_p-+}) hold and \(\mu_{++\atop --}^j\),
\(\mu_{+-\atop -+}^j\) is replaced by \(\hat{\mu}_{++\atop --}^j\),
\(\hat{\mu}_{+-\atop -+}^j\) where in (\ref{mu++--},\ref{mu+--+})
\(\|D'\|\) is replaced by
\(r_j\).
\end{Theorem}
{\bf Proof.} All spectra are real, so we have to find the intersection
of \({\cal C}(\lambda_+^j,\lambda_-^j,r)\) with the real line
the foci \(\lambda_+^j,\lambda_-^j\) from (\ref{Sjlambda})
being also real. This intersection will be a union of two
intervals. For \(\lambda < \lambda_-^j\) and
also for \(\lambda > \lambda_+^j\) the \(j\)-th ovals are given by
\[
(\lambda_-^j - \lambda)(\lambda_+^j - \lambda) \leq -\lambda r
\]
i.e.
\[
\lambda^2 -(\lambda_+^j + \lambda_-^j - r)\lambda +
 \lambda_+^j\lambda_-^j \leq 0
\]
where \(\lambda_+^j + \lambda_-^j = -d_{jj}\) and
\(\lambda_+^j\lambda_-^j = \omega_j^2\). Thus, the left and the right boundary point of the real ovals are \(\mu_{++\atop --}^j\).

For \(\lambda_-^j < \lambda < \lambda_+^j\) the ovals will not
contain \(\lambda\), if
\[
(\lambda - \lambda_-^j)(\lambda_+^j - \lambda) \leq -\lambda r
\]
i.e.
\[
\lambda^2 + (d_{jj} - r)\lambda  + \omega_j^2 < 0
\]
with the solution
\[
\mu_{-+}^j < \lambda < \mu_{+-}^j.
\]
Now take \(r = \|D'\|\). The same argument goes with \(r = r_j\).
Q.E.D.\\

Note the inequality
\begin{equation}\label{mu++--ordered}
(\mu_{--}^j,\mu_{-+}^j) < (\mu_{+-}^k,\mu_{++}^k)
\end{equation}
for all \(j,k\).

{\bf Monotonicity-based bounds.}
As it is known for symmetric matrices monotonicity-based bounds
for the eigenvalues (Wielandt-Hoffmann bounds for a single matrix) have an
important advantage
over Gershgorin-type bounds: While the latter are merely inclusions,
that is, the eigenvalue is contained in a union of intervals
the former tell more: there each interval contains 'its own eigenvalue'.
even if it intersects other intervals.

In this section we will derive bounds of this kind
for overdamped systems.
A basic fact is the following theorem

\begin{Theorem}\label{overd_monot}
With overdamped systems the eigenvalues go asunder
under growing viscosity. More precisely, Let
\[
\lambda_{n-m}^- \leq \cdots \leq \lambda_1^- <
\lambda_1^+ \leq \cdots \leq \lambda_m^+ < 0
\]
be the eigenvalues of an overdamped system \(M,C,K\).
If \(\hat{M},\hat{C},\hat{K}\) is more viscous
that is, \(\hat{M} \leq M,\hat{C} \geq C ,\hat{K} \leq K\)
in the sens of forms then its corresponding
eigenvalues \(\hat{\lambda}_k^\pm\)
satisfy
\begin{equation}\label{eq:overd_monot}
\hat{\lambda}_k^- \leq  \lambda_k^-,\quad
\lambda_k^+\leq  \hat{\lambda}_k^+
\end{equation}
\end{Theorem}
A possible way to prove this theorem is to use the Duffin's
minimax principle \cite{duffin}, moreover, the
following formulae hold
\begin{equation}\label{duffin_minimax}
\lambda_k^+ = \min_{S_k}\max_{x\in S_k}p_+(x),\quad
\lambda_k^- = \max_{S_k}\min_{x\in S_k}p_-(x).
\end{equation}
where \(S_k\) is any \(k\)-dimensional
subspace. Now the proof of Theorem \ref{overd_monot}
is immediate, if we observe that
\begin{equation}\label{p_monot}
\hat{p}_\pm(x) {> \atop <} p_\pm(x)
\end{equation}
for any \(x\).\\

As a natural relative bound for the system matrices
we assume
\begin{equation}\label{MCK_pert}
|x^T\delta Mx| \leq \epsilon x^T Mx,\quad
|x^T\delta Cx| \leq \epsilon x^T Cx,\quad
|x^T\delta Kx| \leq \epsilon x^T Kx,
\end{equation}
with
\begin{equation}\label{MCK_delta}
\delta M = \hat{M} -M,\quad
\delta C = \hat{C} -C,\quad
\delta H = \hat{K} -K,\quad \epsilon < 1.
\end{equation}
We suppose that the system \(M,C,K\) is overdamped
and modally damped.
One readily sees that
the overdampedness of the perturbed system
\(\hat{M},\hat{C},\hat{K}\)  is insured, if
\begin{equation}\label{Delta_pert}
\epsilon < \frac{d - 1}{d + 1},\quad d = \min_x\frac{x^TCx}
{2\sqrt{x^TMxx^TKx}}.
\end{equation}
So, the following three overdamped systems
\[
(1 + \epsilon)M,(1 - \epsilon)C,(1 + \epsilon)K;\quad
\hat{M},\hat{C},\hat{K};\quad
(1 - \epsilon)M,(1 + \epsilon)C,(1 - \epsilon)K
\]
are ordered in growing viscosity. The first and the last
system are overdamped and also modally damped and their eigenvalues
are known and given by
\[
\lambda_k^\pm\left(\frac{1 - \epsilon}{1 +\epsilon}\right),\quad
\lambda_k^\pm\left(\frac{1 + \epsilon}{1 -\epsilon}\right),
\]
respectively, where
\[
\lambda_k^\pm(\eta)
=
\frac{-d_{jj}\eta \pm\sqrt{d_{jj}^2\eta^2 - 4\omega_j^2}}{2},
\]
are the eigenvalues of the
system \(M,\eta C,K\). We suppose that the unperturbed eigenvalues
\(\lambda_k^\pm = \lambda_k^\pm(1)\) are ordered as
\[
\lambda_n^- \leq \cdots \leq \lambda_1^- <
\lambda_1^+ \leq \cdots \leq \lambda_n^+.
\]
By the monotonicity property the corresponding
eigenvalues are bounded as
\begin{equation}\label{l_monot_bound}
\tilde{\lambda}_k^+\left(\frac{1 - \epsilon}{1 +\epsilon}\right)
\leq \hat{\lambda}_k^+ \leq
\tilde{\lambda}_k^+\left(\frac{1 + \epsilon}{1 -\epsilon}\right),
\end{equation}
where \(\tilde{\lambda}_k^+(\eta)\) are obtained by
permuting \(\lambda_k^+(\eta)\) such that
\[
\tilde{\lambda}_1^+(\eta)
\leq \cdots \leq \tilde{\lambda}_n^+(\eta)
\]
for all \(\eta > 0\). It is clear that each \(\tilde{\lambda}_k^+(\eta)\)
is still non-decreasing in \(\eta\). An analogous bound holds for
\(\hat{\lambda}_k^-\) as well.
%



\begin{thebibliography}{ABC-99x}
\bibitem{duffin} Duffin, R.~J., A minimax theory for overdamped networks,
J. Rational Mech. Anal. {\bf 4} (1955),
221--233.
\bibitem{glr} Gohberg, I., Lancaster, P., Rodman, L., Matrices and
   indefinite scalar products, Birkh\"{a}user, Basel 1983.
\bibitem{glr_pol}  Gohberg, I., Lancaster, P., Rodman, L.,
 Matrix polynomials, Academic Press, New York, 1982.
\bibitem{l}  Lancaster, P., Lambda-matrices and vibrating systems,
             Pergamon Press Oxford 1966.
\bibitem{lquad} Lancaster, P., Quadratic eigenvalue problems,
LAA {\bf 150} (1991) 499--506.

\end{thebibliography}
\end{document}